\documentclass{PoS}

\newcommand{\pT}[0]{p_{\mathrm{T}}}

\newcommand{\Nch}[0]{N_{\mathrm{ch}}}

\newcommand{\abs}[1]{\left| #1 \right|} 								
\let\baraccent=\= 													
\renewcommand{\=}[1]{\stackrel{#1}{=}} 								

\newcommand{\figureHeight}{6.5cm}
\usepackage{lineno}
\title{Particle production as a function of system size and underlying-event activity measured with ALICE at the LHC}

\ShortTitle{Particle production with ALICE}

\author{\speaker{Mario Kr\"uger}  for the ALICE Collaboration\\
        Institut f\"ur Kernphysik, Goethe-University Frankfurt\\
        E-mail: \email{mario.kruger@cern.ch}}

\abstract{
		High-energy collisions in ALICE allow for the study of soft and hard QCD particle production.
		The correlation between transverse momentum spectra and event multiplicity is a sensitive observable providing insights into the different production mechanisms. 
		In these proceedings we report on this observable for unidentified charged-particles, obtained using a 2d-unfolding procedure. 
		Particle production is compared at different collision energies, as well as for pp, p--Pb and Pb--Pb collisions at the same energy. In order to understand the role of autocorrelations in small systems, it was proposed to exploit the usage of the underlying event as a multiplicity estimator to factorize the hardest and the softer components of the events. 
		For this purpose, in these proceedings the charged particle transverse momentum distributions are also presented as a function of underlying-event activity in pp collisions.
}

\FullConference{
European Physical Society Conference on High Energy Physics - EPS-HEP2019 -\\
			10-17 July, 2019\\
			Ghent, Belgium}

\begin{document}
\section{Introduction}
Measurements of particle production in heavy-ion collisions at the Large Hadron Collider (LHC) provide insights into a hot and dense deconfined state of matter that was also supposed to have been present shortly after the big bang.
An integral part of the effort to study the properties of this medium is a good understanding of particle production in less complex hadronic collision systems (e.g. pp and p--Pb), where no  Quark--Gluon--Plasma (QGP) is expected to be formed.
It remains still challenging for modern event generators to incorporate the complex interplay of hard and soft physical processes between the strongly interacting partons which ultimately leads to the creation of hadrons in the final state of the collision.
In particular, soft processes cannot be modelled analytically via pertubative approaches and hence their description purely relies on phenomenological models.

Since the energy available in a collision is spent both on the creation of hadrons as well as on their kinetic energy, one fundamental observable to experimentally characterize particle production is the correlation between the amount of particles produced in the collision (multiplicity) and their corresponding transverse momentum ($\pT$) distributions.
Previous measurements of unidentified charged particles by ALICE \cite{meanpt-paper-900} \cite{meanpt-paper} show a rising trend in average $\pT$  with multiplicity for all collision systems at the LHC. This indicates a hardening of the transverse momentum spectra with multiplicity.
In particular for pp collisions it was found that for a proper description of the experimental data it is crucial for the theoretical models to implement the possibility of reconnecting color strings originating from different initial parton scattering centres (multi-parton interactions) \cite{meanpt-paper}.
The analysis presented in these proceedings uses a 2d unfolding procedure to access not only the average $\pT$, but the entire transverse momentum distributions as a function of charged-particle multiplicity ($\Nch$), providing a double differential measurement of the invariant yield with the highest possible granularity in charged-particle multiplicity ($\Delta \Nch = 1$). The results are shown for pp collisions at different energies and for pp, p--Pb and Pb--Pb at the same centre-of-mass energy per nucleon pair.
Complementary to these minimum bias studies it is investigated to what extent jets influence the correlation between multiplicity and $\pT$ in pp collisions at $\sqrt{s}~=$~13~TeV by using a jet-free multiplicity estimator based on the number of charged particles in the underlying event.

\section{Analysis and results}
\begin{figure}[t]
	\centering
	\includegraphics[height=\figureHeight]{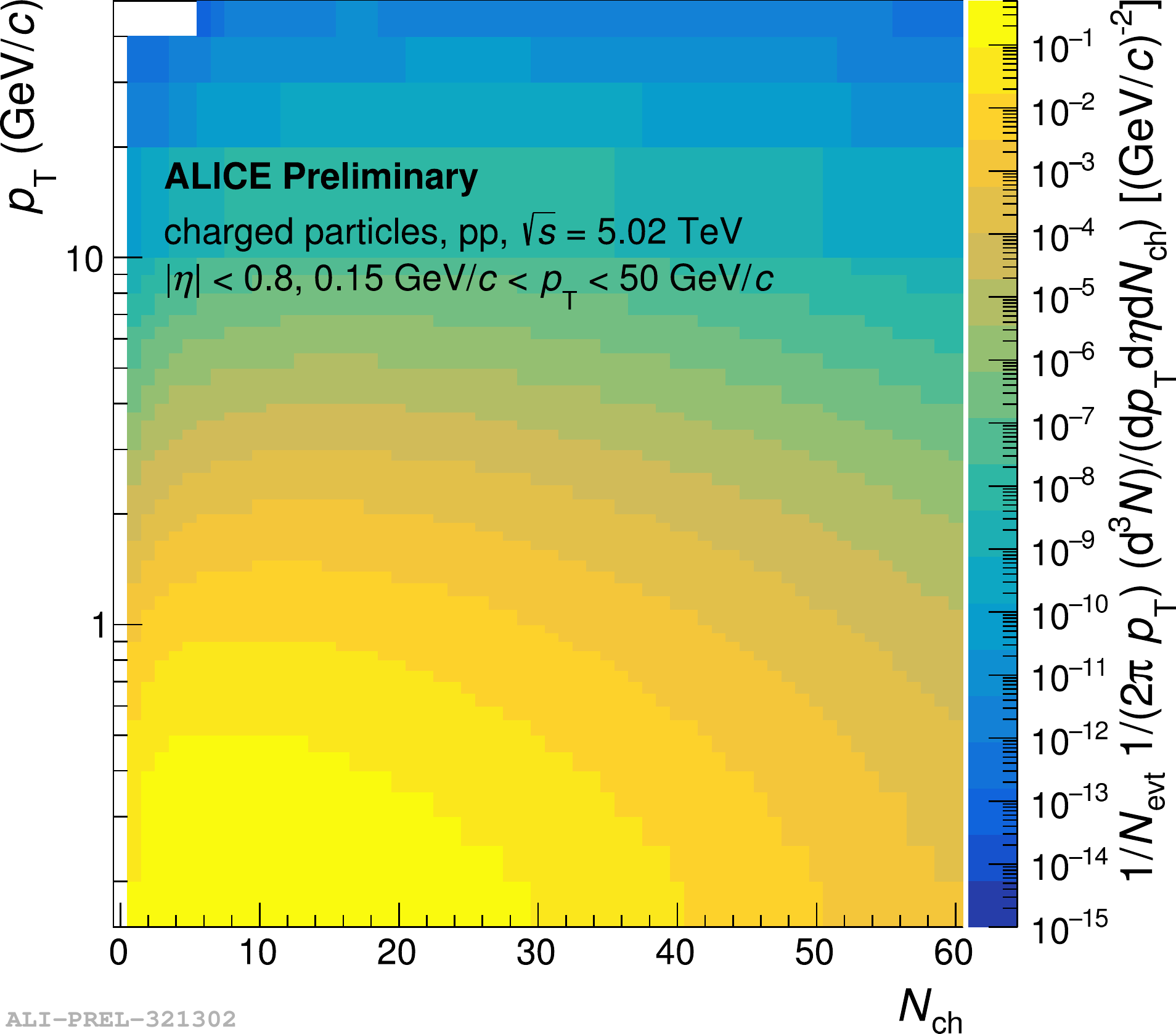} \hfill
	\includegraphics[height=\figureHeight]{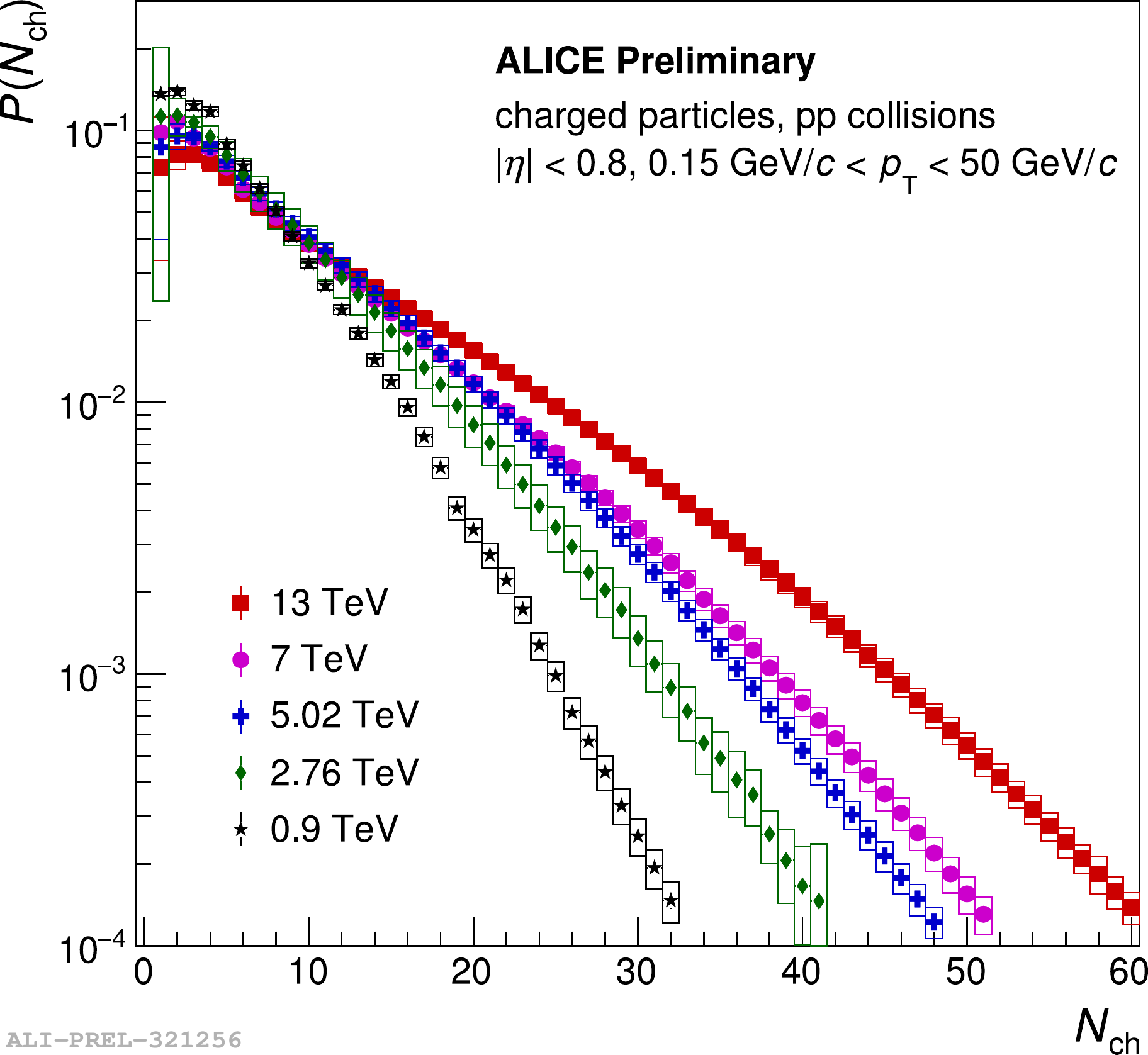}\\
	\vspace{0.5cm}	
	\includegraphics[height=\figureHeight]{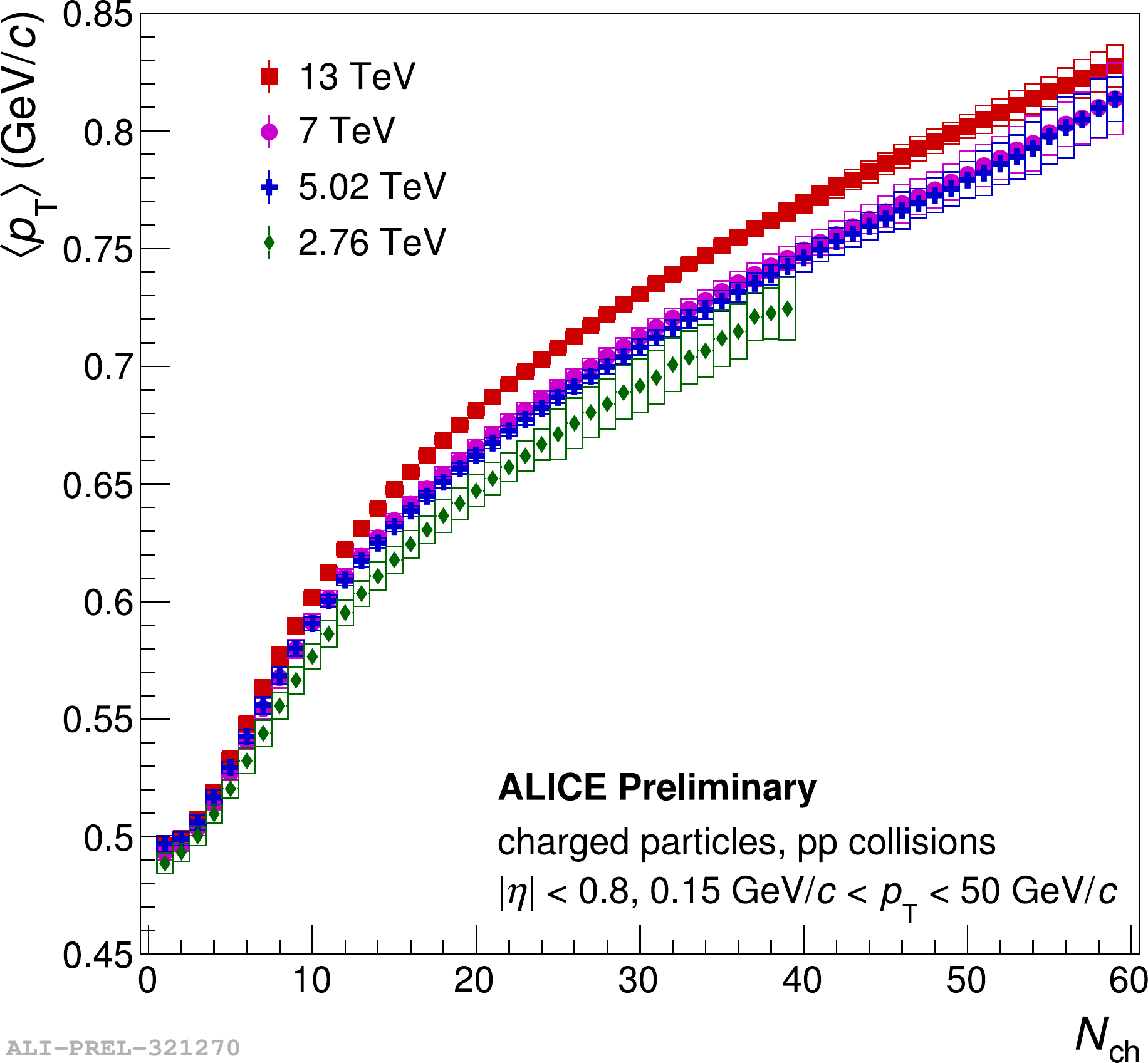} \hfill
	\includegraphics[height=\figureHeight]{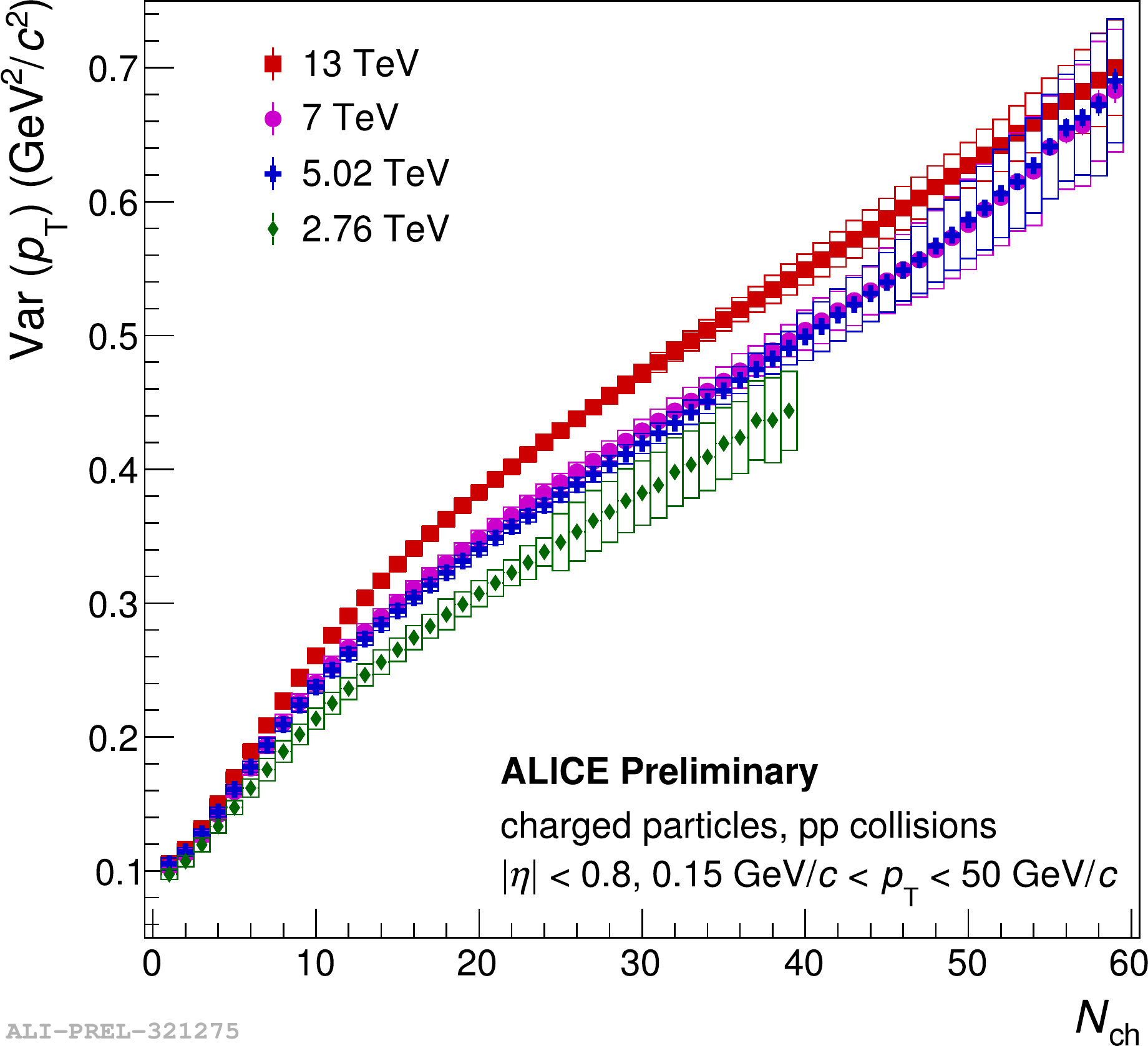} \\
	\vspace{0.5cm}
	\caption{Unfolded multiplicity dependent $\pT$ spectra (upper left panel) and multiplicity distributions (upper right panel). The bottom row shows the mean (left) and variance (right) of the unfolded $\pT$ spectra for pp collisions at different centre-of-mass energies as a function of charged-particle multiplicity.}	
	\label{fig:energyDep}
\end{figure}
\begin{figure}[t]
	\centering
	\includegraphics[height=\figureHeight]{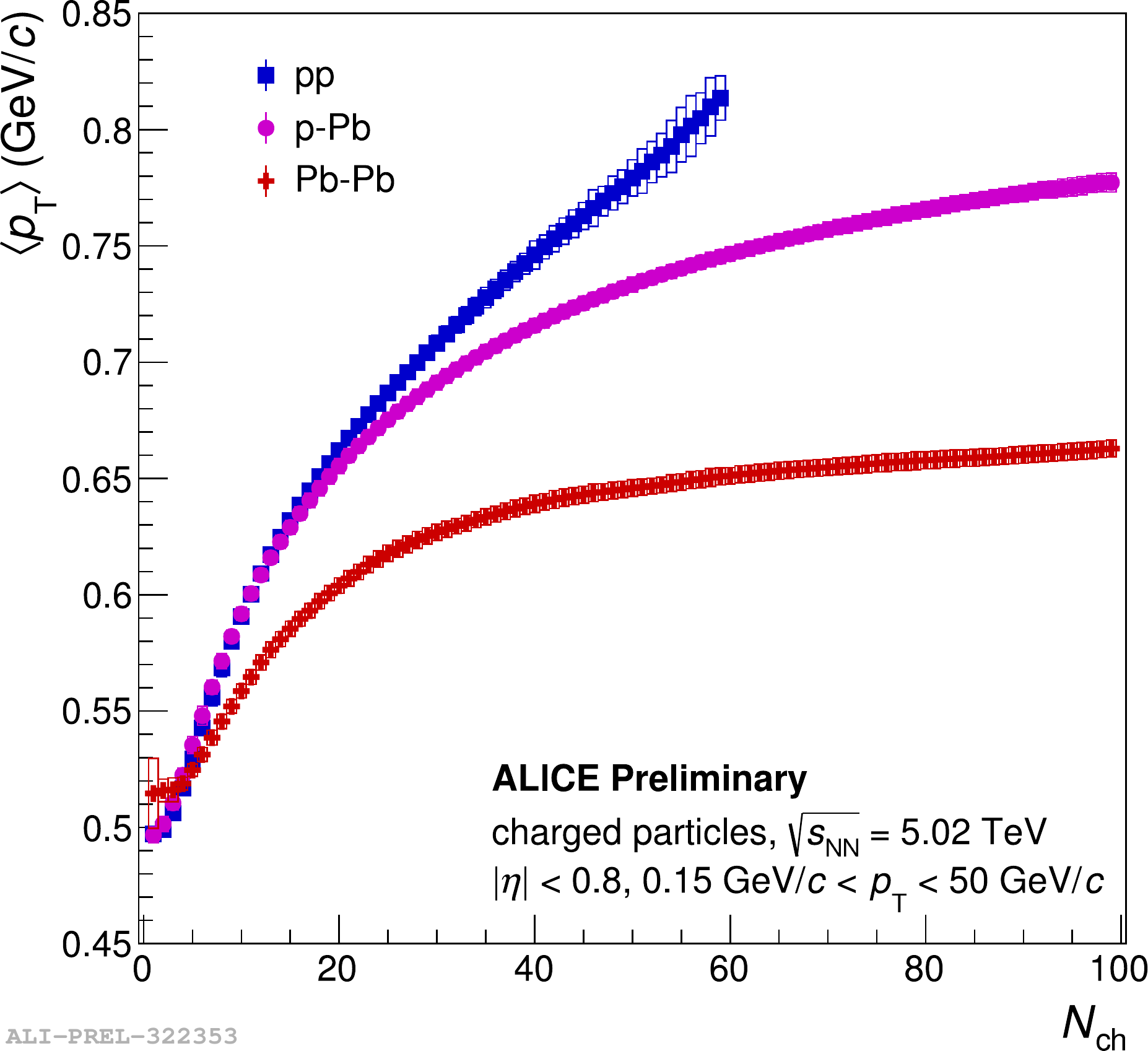}  \hfill
	\includegraphics[height=\figureHeight]{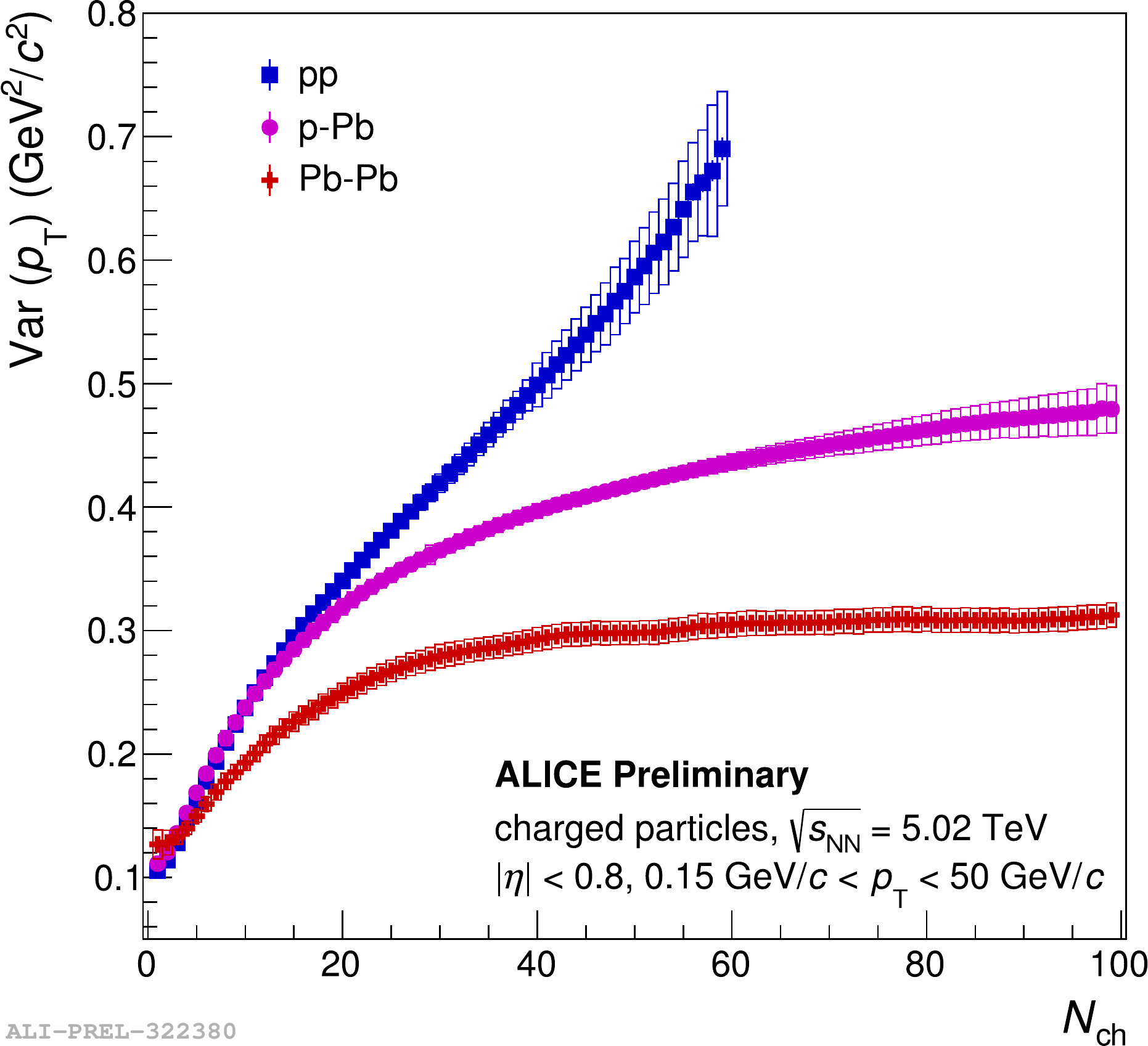}\\
	\vspace{0.5cm}
	\includegraphics[height=\figureHeight]{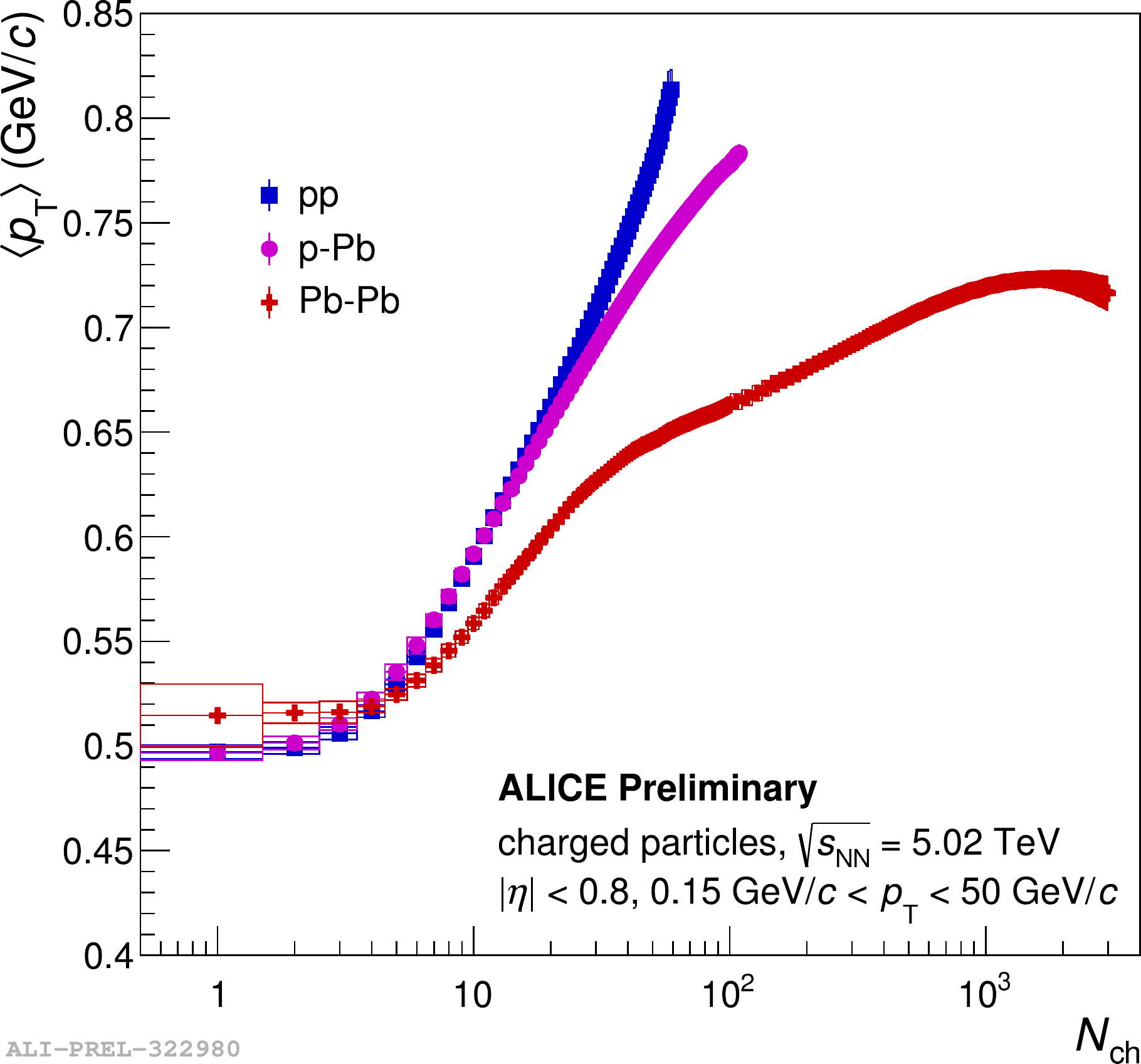}  \hfill
	\includegraphics[height=\figureHeight]{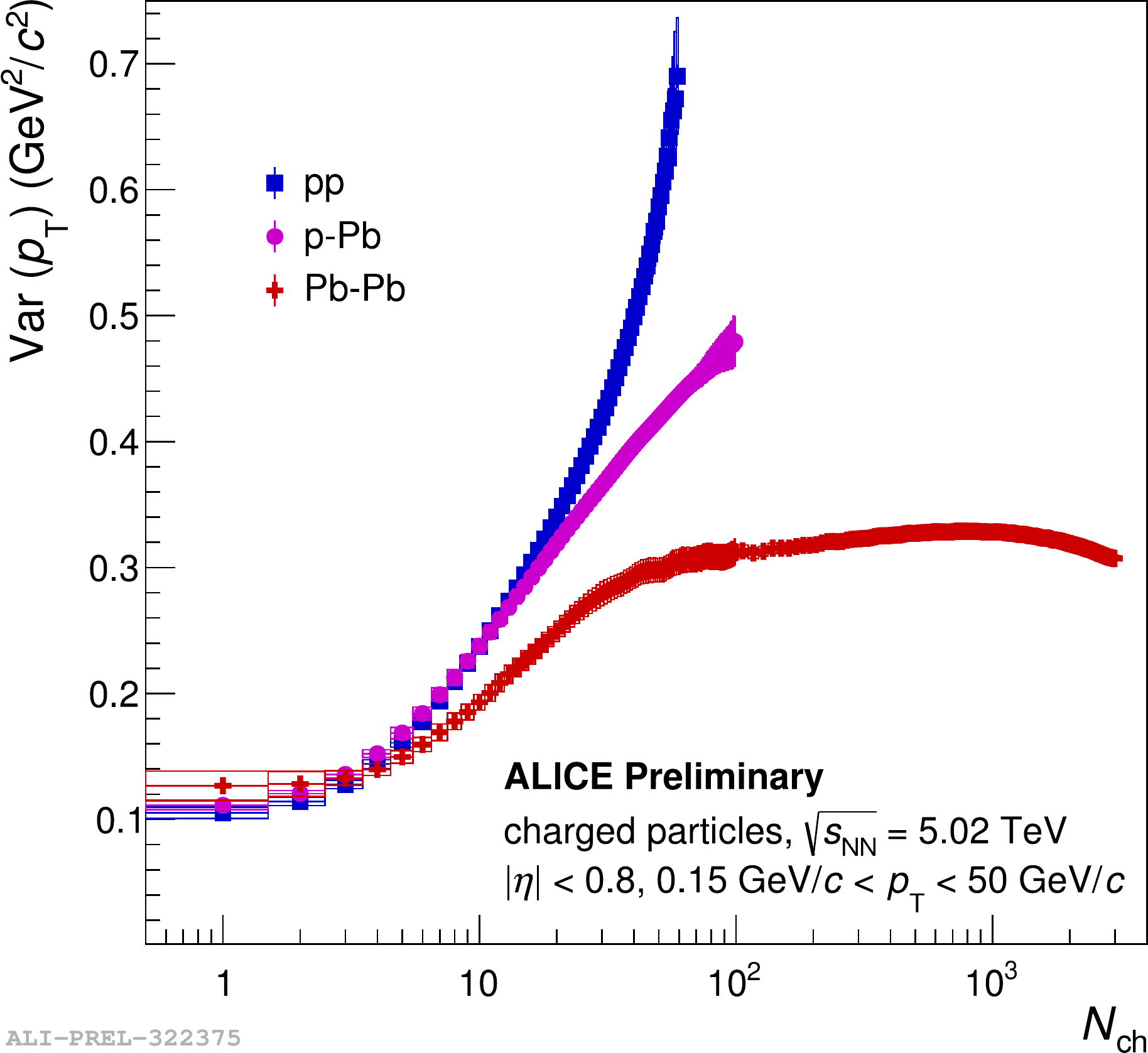} \\
	\vspace{0.5cm}
	\caption{Mean (left panels) and variance (right panels) of the unfolded multiplicity dependent $\pT$ spectra as a function of multiplicity for different system sizes at the same centre-of-mass energy per nucleon pair. The upper panels show the system comparison only for the overlapping multiplicity range with a linear $\Nch$ axis, whilst the lower panels show the full range in multiplicity with a logarithmic scale.}
	\label{fig:systemSizeDep}
\end{figure}
The ALICE apparatus is capable of high-precision measurements of the charged-particle transverse momentum down to very low $\pT$ and as such well suited to study the bulk particle production dominated by soft physics processes.
Within the large solenoid magnet, the tracks of charged particles emerging from the collisions are bent on trajectories that can be reconstructed by combining the 3d space point measurements from the two main tracking detectors: the Inner Tracking System (ITS), a pixel detector close to the vertex, and the large Time Projection Chamber (TPC). A detailed description of the ALICE detectors can be found in \cite{alice}.
To ensure optimal acceptance by the tracking detectors, the vertex position along the LHC beam line is chosen to be $|V_z| < 10 $~cm around the nominal interaction point. Charged particles are selected within a pseudorapidity window of $\abs{\eta} < 0.8$ and with transverse momentum between $0.15~\textrm{GeV}/c$ and  $ 50~\textrm{GeV}/c$.
The data are supplemented by Monte Carlo simulations consisting of an event generator and a virtual GEANT \cite{geant} model of the detector, which gives access to performance and acceptance effects.
Event-by-event fluctuations of the detector efficiency result in a smearing of the multiplicity measurement and therefore also of the correlation between $\Nch$ and $\pT$. This detector effect can be corrected for by means of an iterative unfolding method \cite{agostini-bayes} \cite{proceedingsQCHS} which is used to  un-smear the multiplicity measurement as well as the corresponding $\pT$ distributions.
Corrections for efficiency and acceptance losses as well as the residual contamination with secondary particles are included in the unfolding.
A simulation-based study of the performance of the method can be found at \cite{proceedingsQCHS}.

In the upper left panel of Figure \ref{fig:energyDep} the resulting double-differential invariant yield of charged particles obtained with this procedure is shown for pp collisions at $\sqrt{s} =$ 5.02 TeV.
From the unfolded multiplicity distributions in the upper right panel of Figure \ref{fig:energyDep} one can see that the reach in $\Nch$ strongly depends on the overall centre-of-mass energy available in the collision.
Differences of the unfolded multiplicity dependent $\pT$ spectra can be illustrated by comparing simple derived quantities that characterize the spectral shape, e.g. the mean or the variance of the different  transverse momentum distributions.
In Figure \ref{fig:energyDep}, the mean $\pT$ (bottom left panel) and variance (bottom right panel) are shown for pp collisions at $\sqrt{s} = $ 2.76 TeV, 5.02 TeV, 7 TeV and 13 TeV. 
Notably the $\langle\pT\rangle$ is the same ($ \langle\pT\rangle \approx 0.49~\mathrm{GeV}/c$ ) at low multiplicities  for all the collision energies, for higher $N_{\mathrm{ch}}$ the distributions deviate ordered by energy.
A similar observation can be made for the variance of the transverse momentum distributions displayed in the lower right panel.
In Figure \ref{fig:systemSizeDep}, the mean $\pT$ and variance are shown for pp p--Pb and Pb--Pb collisions at the same centre-of-mass energy per nucleon pair of $\sqrt{s_{\mathrm{NN}}}~= $~5.02 TeV. 
The rise of the mean $\pT$ and variance is steepest for pp collisions whilst it is least pronounced in Pb--Pb, where according to models the particles produced in the collision are re-scattered among the nucleons.
In p--Pb a hybrid behaviour is observed.
The p--Pb $\pT$ spectra are very similar to the spectra in pp up to $N_{\mathrm{ch}} \approx 15-20$. For higher $N_{\mathrm{ch}}$ the slope of mean $\pT$ and variance decreases, analogously to what is observed in Pb--Pb collisions.
Since the plots shown with linear scale in the upper panels of Figure \ref{fig:systemSizeDep} only represent very peripheral Pb--Pb events  (centrality $\gtrsim 70\%$), the lower panels of this figure additionally show the whole Pb--Pb multiplicity range with a logarithmic scale.
Notably, the maximum mean $\pT$ and variance measured in Pb--Pb collisions are smaller than the maximum of these quantities observed in pp or p--Pb, which indicates an influence of the presence of the Pb nucleons in the final state of the collision.
Additionally, a decrease in $\langle\pT\rangle$ and variance is observed for very high multiplicities, which correspond to the most central Pb--Pb collisions.
\begin{figure}[t]
	\centering
	\includegraphics[height=\figureHeight]{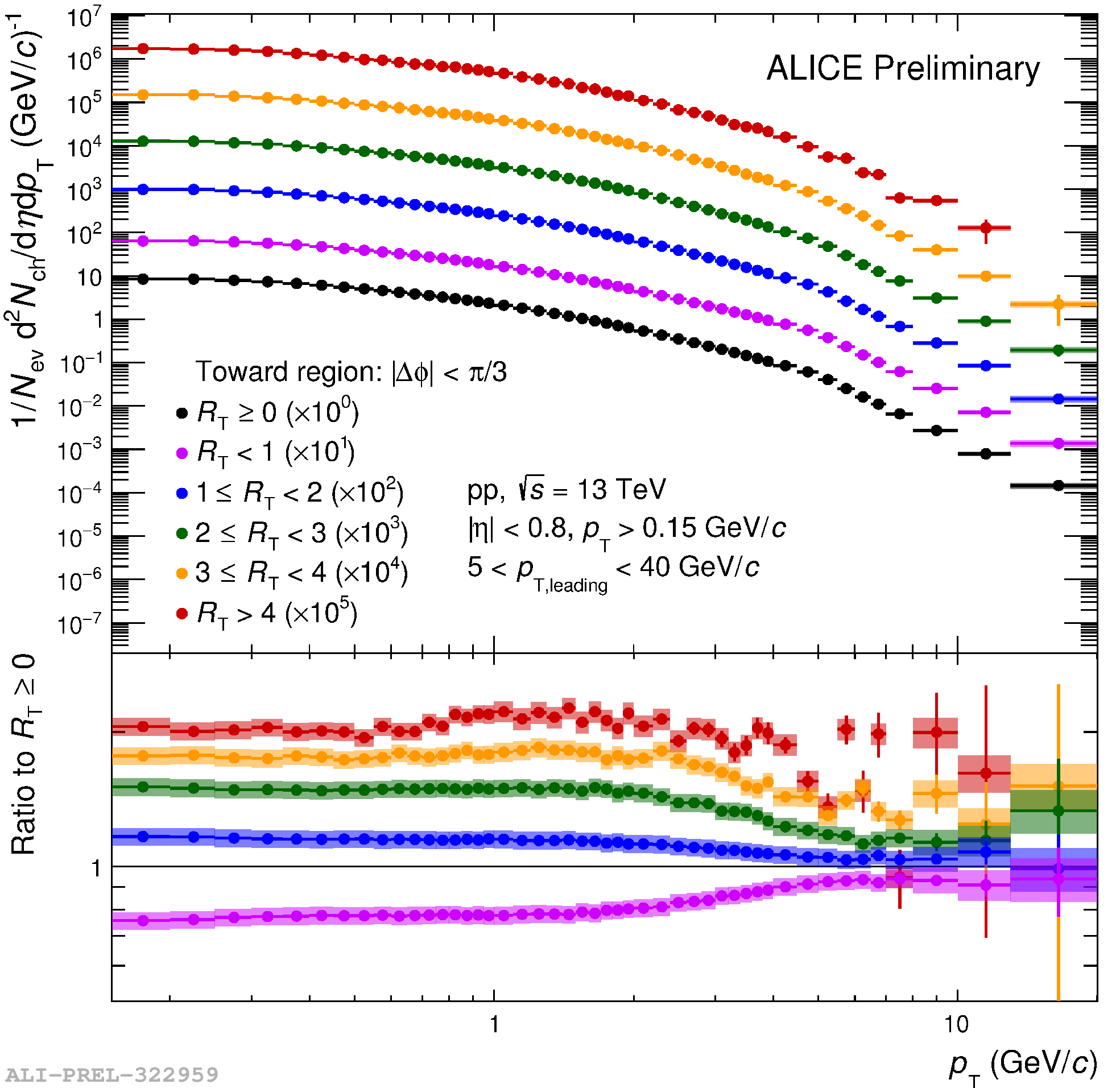} \hfill
	\includegraphics[height=\figureHeight]{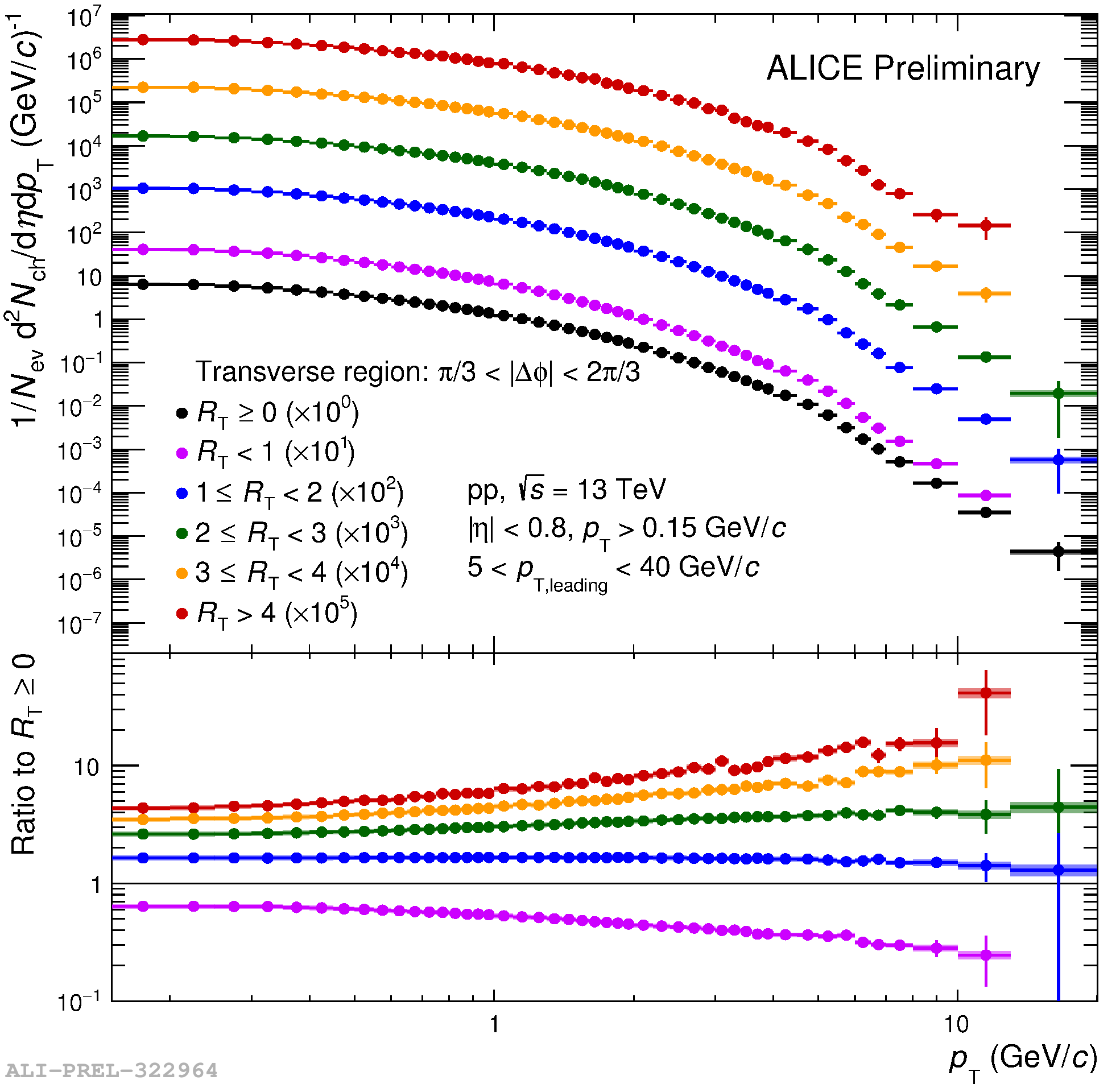} 
	\caption{Transverse momentum spectra classified by underlying event activity
		 $R_{\mathrm{T}}$ as measured in the region along the jet (left panel) and in the region transverse to the jet axis (right panel).}
	\label{fig:underlyingEvent}
\end{figure}

In all results previously shown the multiplicities and their corresponding transverse momentum spectra are consistently measured in the same kinematic range, which means that particles contained in the spectra are also defining the respective multiplicities.
The correlation between $\Nch$ and $\pT$ therefore represents the whole complexity of all the hard and soft physical processes leading to particle production.
In particular, particle jets strongly influence the correlation between multiplicity and $\pT$ as they generally consist of many particles with rather high transverse momentum.
One approach to factorize hard and soft particle production in pp collisions is to explicitly select only particles not originating from the hardest initial scattering that usually result in the formation of a jet.
It was proposed \cite{Martin:2016igp} to facilitate the so-called underlying event to benchmark phenomenological models implemented in modern event generators.
As an experimental definition of the underlying event, here the region transverse to the leading jet in a collision is used.
The direction of this leading jet is approximated by a cone ($\phi  \pm 30^\circ$) around the charged-particle track with the highest  $\pT$.
Assuming that the region transverse to this jet cone ($30^\circ < \abs{\phi} < 120^\circ$) is free of remnants from the original back-to-back particle jets created in the collision, the underlying event activity $R_{\mathrm{T}}$ can be defined as the charged-particle multiplicity $N_{\mathrm{ch, T}}$ in this transverse region with respect to its event averaged mean:
\begin{equation}
	R_{\mathrm{T}} = \frac{N_{\mathrm{ch, T}}}{\langle N_{\mathrm{ch, T}} \rangle}
\end{equation}
Figure \ref{fig:underlyingEvent} shows the measured charged particle $\pT$ spectra as a function of $R_{\mathrm{T}}$ for pp collisions at $\sqrt{s} = 13 $ TeV.
The black markers represent the $R_{\mathrm{T}}$ integrated spectra and the coloured markers five different classes of underlying event activity.
In the left panel of Figure \ref{fig:underlyingEvent} the $\pT$ spectra measured alongside the jet (toward region) are shown.
The ratios of the $R_{\mathrm{T}}$ selected spectra to the inclusive spectrum displayed in the lower part of the plot illustrate that at high transverse momentum the measured particle yield hardly depends on the underlying event activity.
This indicates that along the jet at high $\pT$ there is a clear separation between hard and soft particle production.
The $R_{\mathrm{T}}$ dependent $\pT$ spectra measured in the region transverse to the jet, are shown in the right panel of Figure \ref{fig:underlyingEvent}.
Even though here the hardest jet of the collision is excluded in both the spectra and the multiplicity, the ratio shown in the bottom part of the plot reveals a hardening of the $\pT$ spectra with $R_{\mathrm{T}}$, similar to the results obtained for the minimum bias measurements shown in these proceedings and in \cite{spherocityPaper}. This might be a remaining auto-correlation effect caused by semi-hard processes contained in the underlying event.

\bibliographystyle{JHEP}
\bibliography{Sources}

\end{document}